\documentclass[
    ,final            
  ]
  {aipproc}

\usepackage{amsmath}

\layoutstyle{8x11double}

\begin{document}

\newcommand{\tfe}{1E 1048.1$-$5937}
\newcommand{\ssec}{\textrm{s}}

\title{Red Noise in Anomalous X-ray Pulsar Timing Residuals}
\author{Anne M. Archibald}{
  address={Department of Physics, McGill University, 3600 University Street, Montreal, QC, H3A 2T8, Canada}
}
\author{Rim Dib}{
  address={Department of Physics, McGill University, 3600 University Street, Montreal, QC, H3A 2T8, Canada}
}
\author{Margaret A. Livingstone}{
  address={Department of Physics, McGill University, 3600 University Street, Montreal, QC, H3A 2T8, Canada}
}
\author{Victoria M. Kaspi}{
  address={Department of Physics, McGill University, 3600 University Street, Montreal, QC, H3A 2T8, Canada}
}

\begin{abstract}
Anomalous X-ray Pulsars (AXPs), thought to be magnetars, exhibit poorly understood deviations from a simple spin-down called ``timing noise''.  AXP timing noise has strong low-frequency components which pose significant challenges for quantification. We describe a procedure for extracting two quantities of interest, the intensity and power spectral index of timing noise. We apply this procedure to timing data from three sources: a monitoring campaign of five AXPs, observations of five young pulsars, and the stable rotator PSR B1937+21.
\end{abstract}

\maketitle

\section{Introduction}
\begin{figure}
\includegraphics[width=\columnwidth]{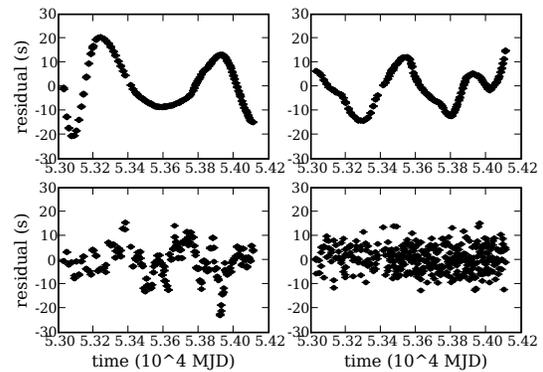}
\caption{Timing noise. Observations of the AXP \tfe\ (top left) are compared with algorithmically generated timing noise with power spectral indices of $-4$ (top right), $-2$ (bottom left) and $0$ (bottom right). In all cases the best-fit cubic has been subtracted (accounting for two derivatives of frequency). Observe that timing noise with a large negative spectral index takes on a smooth appearance because of the strong low-frequency components.}
\label{fig:noise-shapes}
\end{figure}

Most pulsars spin with extreme regularity. Some have spin-down stability comparable to the best atomic clocks. Others, however, appear to deviate from a simple spin-down, requiring a polynomial of high degree to model their rotation over even a year (see Figure~\ref{fig:noise-shapes} for some examples). Still others are sufficiently unpredictable that phase-connecting the observations---that is, exactly counting the number of revolutions between observations---is difficult or impossible. The causes of such deviations, called ``timing noise'', remain poorly understood \citep{hlk06}.

Anomalous X-ray pulsars (AXPs) are slow pulsars (periods 2--12 s) whose X-ray luminosity exceeds the power available from spin-down. The leading description of these objects is the magnetar model \citep{tlk02}. Eight AXPs are known, plus two candidates.\footnote{An online catalog of AXPs and SGRs is maintained by Cindy Tam at \url{http://www.physics.mcgill.ca/~pulsar/magnetar/main.html}.} AXPs exhibit a tremendous range of variability in their radiative behaviour, including flux and profile variations (\cite{t+07}, \citep{dkg07}), bursts \citep{gkw04}, glitches \citep{dkg07b}, and strong timing noise. In fact, the AXP \tfe\ has intervals during which phase connection is difficult to impossible and periods where it is straightforward (in spite of an unchanged cadence of observations). This indicates that the nature or amplitude of its timing noise changes. By quantifying the timing noise, we hope not only to describe these changes but to probe the differences between AXPs and other classes of pulsar.

\section{Analysis}
The analysis of timing noise poses a number of technical difficulties. Observed timing noise has a steep red spectrum, with power spectral indices often near $-4$ (see Table~\ref{tab:results}). Noise with such a red spectrum cannot be analyzed with normal Fourier techniques, as individual Fourier bins do not reject low frequencies strongly enough (this is often called ``spectral bleeding'') \citep{db82}. Further, the span of data limits spectral resolution at low frequencies no matter what technique is used. 

\begin{figure}
\includegraphics[width=\columnwidth]{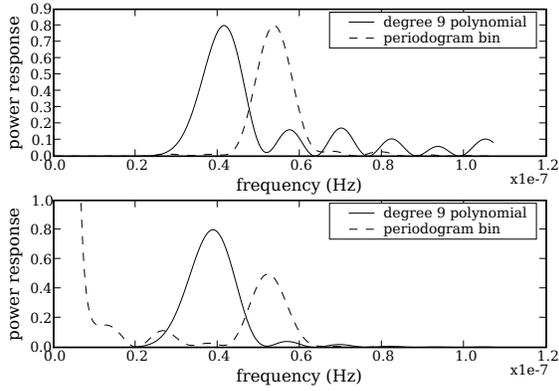}
\caption{Spectral responses of orthogonal polynomials and periodogram bins, sampled at the times of our observations of \tfe. The top plot compares the response of a degree 9 orthogonal polynomial and a periodogram bin to a sine wave. The bottom plot shows the same response scaled by $f^{-4}$. This indicates which parts of a red spectrum the filters respond to. Observe that the periodogram is much more sensitive to the lowest frequencies in such a red spectrum than it is to the frequencies at its nominal peak, while the orthogonal polynomial rejects them strongly.}
\label{fig:spectral-responses}
\end{figure}

To bypass the limitations of Fourier analysis, we are developing a method that we believe will improve spectral analysis of red noise. Our code constructs a filterbank for each data set based on orthogonal polynomials, extending the procedure described by \citet{db82}. Specifically, \citet{db82} describe a procedure for computing a family of polynomials orthogonal with respect to the white observational noise in a data set. The resulting filters reject low frequencies very strongly (see Figure~\ref{fig:spectral-responses}). 
They then carry out a second orthogonalization procedure based on a particular red noise model, and they select a single filter from the resulting filterbank.To examine higher frequencies, they choose to apply this process from scratch on halves, quarters, eighths, etc. of the original time series. 

By comparison, we compute a family of polynomials orthogonal with respect to the white noise, as \citet{db82} do. However, we simply use these polynomials as our filterbank; their frequency responses vary, but as a filterbank they provide reasonable frequency resolution at high frequencies, and strong rejection of low frequencies. The polynomials remain covariant with respect to timing noise, so we take that into account in our fitting procedure. Given a power-law model of timing noise (as described in \citet{sfw03}), we can compute the likelihood of the observed filter values using the filterbank's covariance matrix with respect to this red noise model. We then carry out a computationally expensive numerical optimization procedure to find the maximum-likelihood power-law model.

The power-law spectral model we fit to the timing noise is
\begin{equation}
F(\omega) = A\left(\frac{\omega}{10^{-8} \text{Hz}}\right)^\alpha,
\end{equation}
where $F$ is the power spectral density in $s^2/\text{Hz}$ and $\omega$ is the frequency of the timing noise. Fitting then gives us $A$ and $\alpha$. We also take into account the uncertainties on our arrival-time measurements, which are independent. This white timing noise, due to observational effects, sets an upper limit on the frequencies we can include in our model.
\begin{figure}
\includegraphics[width=\columnwidth]{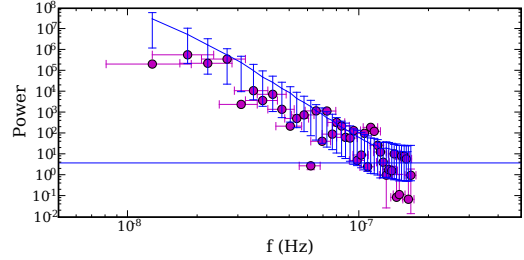}
\caption{A model fit to observations of \tfe. The circles are the amplitudes of the responses of an orthogonal filterbank to the residuals pictured in Figure~\ref{fig:noise-shapes}. The horizontal error bars indicate the full width at half-maximum response of each filter. If the best-fit model we computed is the true model, we expect the amplitude returned by each filter to be distributed according to a $\chi^2$ distribution with one degree of freedom. The line connects the expected value for each such distribution, and the vertical error bars attached to it indicate the $\pm 1 \sigma$ points. The horizontal line indicates the background level of white noise (coming mostly from observational uncertainties); this is also indicated with vertical error bars attached to the circles. For most of the plotted filters the observational uncertainty is negligible compared to the variability we expect from the timing noise.}
\label{fig:fit}
\end{figure}

\begin{table}
\begin{tabular}{lcccc}
\hline
\tablehead{1}{l}{c}{Source} & 
\tablehead{1}{c}{c}{Exponent } &
\tablehead{1}{c}{c}{log amplitude ($\ssec^2/\text{Hz}$)} & 
\tablehead{1}{c}{c}{Period ($\ssec$)} & 
\tablehead{1}{c}{c}{Dates (MJD)}
\\
\hline
\tfe & $-4.3\pm 0.2$ & $10.8\pm 0.3$ & 6.5 & 53030--54111 \\
1E 1841$-$045 & $-4.3\pm 0.9$ & $8.9\pm 0.4$ & 11.8 & 51225--52438 \\
              & $-4.3\pm 1.1$ & $9.1\pm 0.7$ &      & 53030--53816 \\
1RXS J170849.0$-$400910 & $-5.7\pm 2.3$ & $8.1\pm 2.3$ & 11.0 & 53638--54016 \\
4U 0142+61 & $-4.4\pm 1.3$ & $6.3\pm 0.4$ & 8.7 & 51611--53787 \\
1E 2259+586\footnote{The timing noise model is a poor fit to the data for this source} & $-2\pm 2$ & $4.7\pm 0.9$ & 7.0 & 53012--54181  \\
\hline
PSR B0540$-$69 & $-4.7\pm 0.8$ & $3.2\pm 0.4$ & 0.050 & 51342--52936 \\
AX J1811.5$-$2134 & $-3.7\pm 1.3$ & $3.4\pm 0.6$ & 0.065 & 52342--53333 \\
PSR B1937+21 & $-3.6\pm 1.2$ & $-6.6\pm 0.3$ & 0.0016 & 45986--49303 \\
\hline
\end{tabular}
\label{tab:results}
\caption{Preliminary timing noise model fits. Each row represents a single span of data; separate spans for the same source were fit independently.}
\end{table}

\section{Data}

We are considering monitoring data for three kinds of sources: five AXPs, five young pulsars, and the stable rotator PSR B1937+21.

Of the known AXPs, we consider monitoring data five (\tfe, 1E 2259+586, 4U 0142+61, 1E 1841$-$045, and RXS J170849.0$-$400910) on a regular basis using the Rossi X-ray Timing Explorer (\emph{RXTE}), as described in Dib et al. 2007 (this volume). Observations, spanning a decade, were generally carried out on a weekly or monthly basis, and yielded time-of-arrival uncertainties generally less than 1\% of a period. These data sets have been phase-connected where possible. Glitches have been observed in all these AXPs.  We consider only inter-glitch intervals, as it is frequently impossible to unambiguously count phases across a glitch, and because glitches seem likely to be produced by a different mechanism than timing noise \citep{cg81}. We were also forced to exclude intervals where the pulse profile changed significantly (making identification of a fiducial point difficult), intervals where phase-connecting the observations was impossible, and intervals where the span of good data was too short to be useful. 


We also consider a collection of observations (primarily using \emph{RXTE}) of five young pulsars. In particular:  we consider 7.6 years of observations (interrupted by a glitch) of PSR B0540$-$69 \citep{lkg05}; we consider 2.7 years of observations of PSR J1811$-$1925 (in the supernova remnant (SNR) G11.2$-$0.3); we consider 7 years of observations (interrupted by a data gap) of PSR J1846$-$0258 (in the SNR Kes 75) \citep{l+06}; we consider 2.7 years of combined X-ray and radio (using the Robert C. Byrd telescope at the Green Bank observatory and the Lovell telescope at Jodrell Bank) observations of PSR J0205+6449 (in the SNR 3C 58) (Livingstone et al. 2007, these proceedings); and we consider 21.3 years of combined X-ray and radio observations (with the Molonglo Observatory Synthesis Telescope and Parkes) of PSR B1509$-$58 \citep{l+05}.

The pulsar PSR B1937+21 is a millisecond pulsar with exceptionally stable rotation. For this reason its rotation has been monitored carefully (see, for example, \citet{ktr94}), and we consider 9.0 years of radio observations obtained with the radio telescope at Arecibo.

\section{Results and Discussion}
Our code still requires work. Nevertheless we can give some preliminary measurements that make quantitative the claim that AXPs have very strong timing noise; see Table~\ref{tab:results}. Observe that the AXP timing noise is, for the most part, vastly stronger than the timing noise in the other sources, and although all the exponents are consistent with $-4$ (the value for white torque noise) there is a hint that AXPs may have slightly redder timing noise than other pulsars. 

We hope to be able to use our code and data to answer several questions. Do AXPs have time-varying timing noise? Does the timing noise in AXPs differ in character from the timing noise in other classes of pulsar? Can we estimate the degree to which measurements of proper motions, braking indices, and so on are contaminated by timing noise? We also hope eventually to extend our analysis to a broader range of radio pulsars, and possibly also to Soft Gamma Repeaters, if this is feasible.

\bibliographystyle{aipproc}
\bibliography{journals_apj,refs}

\end{document}